\documentclass[runningheads]{apinv}
\usepackage{epsfig,graphics}
\usepackage{marvosym} 
\usepackage{natbib}

\usepackage[utf8]{inputenc}

\usepackage[pagewise]{lineno}

\begin{document}

\title{Symbiotic stars in Galactic open clusters}
\titlerunning{Symbiotic stars in Galactic open clusters}
\author{Lukas Kueß\inst{1}, Ernst Paunzen\inst{2}}
\authorrunning{L.Kueß, E.Paunzen}
\tocauthor{Lukas Kueß, Ernst Paunzen} 
\institute{Department of Astrophysics, Vienna University, T{\"u}rkenschanzstraße 17, 1180 Vienna, Austria
        \newline
        \email{lukaskuess@gmx.at}
	\and Department of Theoretical Physics and Astrophysics, Faculty of Science, Masaryk University, Kotl\'{a}\v{r}sk\'{a} 2, 611\,37 Brno, Czechia  \newline
	\email{epaunzen@physics.muni.cz}    }
\papertype{Submitted on xx.xx.xxxx; Accepted on xx.xx.xxxx}	
\maketitle

\begin{abstract}
The age determination of symbiotic stars is essential to put further constraints on models explaining 
these binary systems. In the Galactic field, this is especially problematic because of 
several limitations due to reddening estimations, for example. We searched for symbiotic stars as 
members of Galactic open clusters for which the age and overall metallicity can be determined in a 
statistical sense. The most recent lists of well-established and candidate symbiotic stars
and open clusters were matched, and
we found seven good candidates from which the well-established symbiotic 
star CQ Dra seems to be a true member of the old open cluster HSC 1224. The 
colour-magnitude diagrams for the other candidates raise some doubts about membership. 
\end{abstract}

\section*{Introduction}

Symbiotic stars are a particular type of interacting binaries. They generally consists of three components: (i) a hot star, typically a white dwarf or neutron star with an accretion disk, (ii) a cool star, either a red giant or a star on the asymptotic giant branch (AGB), and (iii) a nebula consisting of ionized material that has been lost by the cool component \citep{2019arXiv190901389M}. The nebulae surrounding these objects are detected via various emission lines.\\
Symbiotic stars represent a crusial stage in the binary evolution of low and intermediate-mass stars. Among other objects, they also are candidate progenitors of supernovae of type Ia \citep[SNe Ia; ][]{2013IAUS..281..162M} Recently, \citet{2025A&A...698A.155L} argued that the contribution of the single-degenerate channel of SNe Ia from symbiotic progenitors is estimated to be on the order of 1\% for our galaxy. Additionally, symbiotic stars are important sources of soft and hard X-rays \citep{2013A&A...559A...6L, 1997A&A...319..201M}.

These objects provide insights for studying the loss of matter, acceleration mechanisms of
stellar winds, and accretion of stellar winds in late-type giants \citep{2019A&A...626A..68S}.
They also allow the analysis of mass transfer, the characteristics of accretion-disk boundary
layers, and astrophysical jets \citep{2017A&A...602A..53S}. 

Symbiotic stars are classified into two main categories based on their
near-infrared data \citep{1975MNRAS.171..171W}: (i) those with a near-IR colour temperature 
of a K, M, or G-type giant (3000 to 4000\,K; stellar or S-type), and
(ii) those with a near-IR colour temperature of around 700 to 1000\,K,
indicating a circumstellar envelope surrounding a more evolved AGB
star. Later on, further subgroups were defined or suggested 
by \citet{1982ASSL...95...27A,1987A&A...182...51N,2019ApJS..240...21A}.

However, a lot of questions remain unanswered, probably because the number of confirmed
Galactic symbiotic stars is about nearly 300, with about 750 additional being most likely 
or suspected ones \footnote{\url{https://sirrah.troja.mff.cuni.cz/\textasciitilde merc/nodsv/}}. 
A lot of efforts were put into unambiguously 
commonly identifying characteristics, using photometry and spectroscopy over the 
entire electromagnetic spectrum 
\citep{2019ApJS..240...21A,2023MNRAS.519.6044A,2021MNRAS.505.6121M,2021MNRAS.506.4151M,2024arXiv241200855L}. 
Because of the typical binary separations of a few to tens of AU,
they observed orbital periods on the order of hundreds of days to a few years
\citep{2013AcA....63..405G}. 

The age and lifespan of symbiotic stars mainly depend on the mass of the cool 
component. A star with 8\,M$_\odot$ on the main sequence evolves in about 35\,Myr to the
red giant phase. It stays there for just a very short time. If we exclude the possibility that
blue giants and supergiants may also form a corresponding binary system, about 35\,Myr 
(depending on the metallicity and rotation) seems the lower limit for the age of
symbiotic stars.

As described, symbiotic stars' evolutionary status is essential to constrain 
the corresponding evolutionary models. It is well known that the age determination of 
Galactic field stars is not straightforward. For cool type stars, it can be done via
isochrone fitting \citep{2021A&A...649A.127V}, gyro-kinematic 
techniques \citep{2024MNRAS.533.1290S} 
or asteroseismology \citep{2024A&A...684A.112F}, to mention a few. Because symbiotic stars
are rather ''exotic`` binaries with unusual reddening values and energy distributions, most
previously mentioned methods will result in large errors.

\begin{figure}[!tb]
    \centering
    \includegraphics[width=\textwidth]{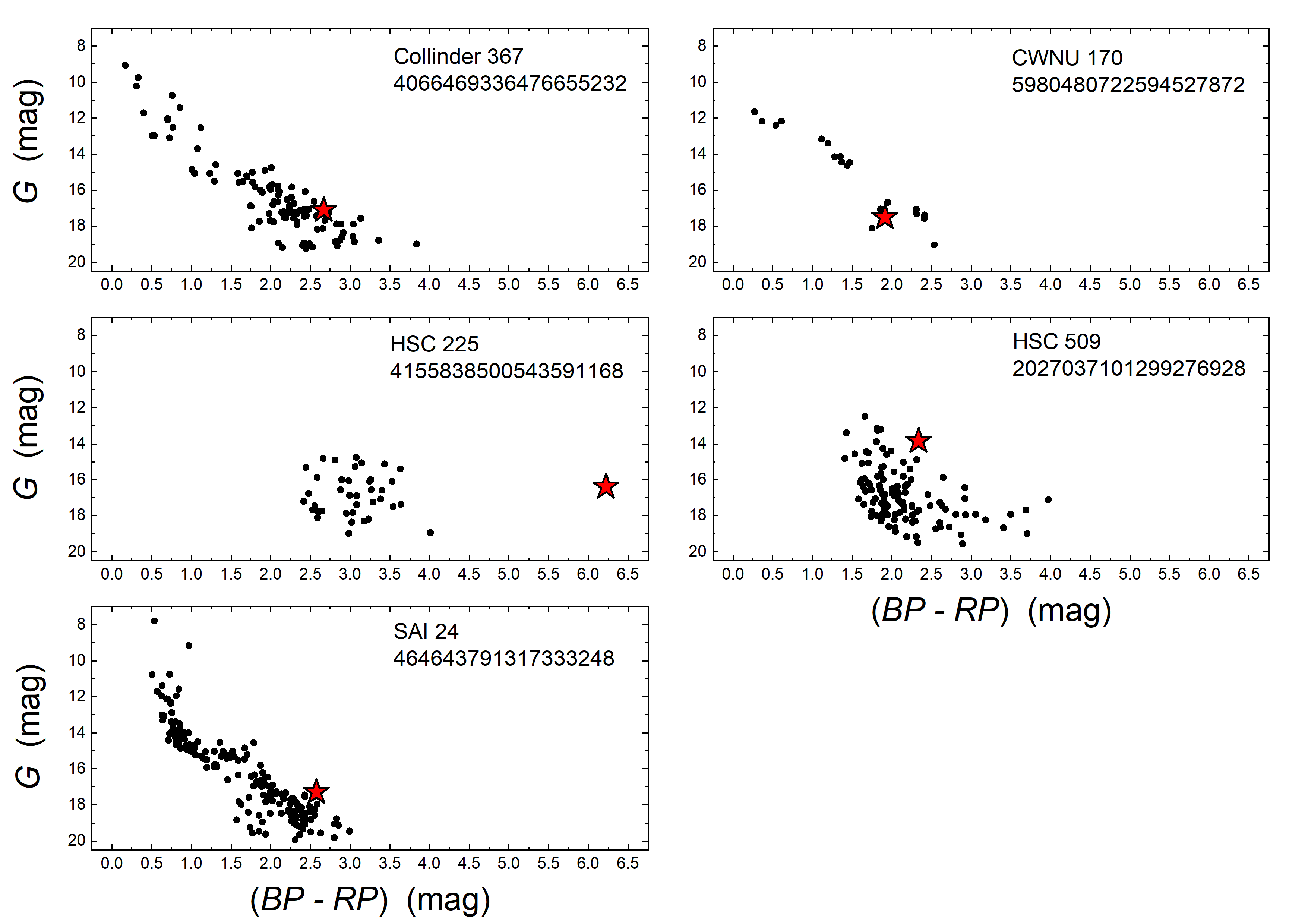}
    \caption{The CMDs of the open clusters Collinder 367, CWNU 170, HSC 225, HSC 509, and SAI 24 together with the corresponding symbiotic candidate.}
    \label{fig:others}
\end{figure}

\begin{table*}[!tb]
\begin{center}
\caption{Galactic symbiotic stars taken from the latest version of the catalogue by
\citet{2019RNAAS...3...28M} as possible members of star clusters. The columns ``Name'' and ``Class'' are taken from the aforementioned list.}
\label{table:canidates}
\tiny
\begin{tabular}{cccccc}
$Gaia$ DR3\_ID & Name & Class & $\alpha$ (2000) & $\delta$ (2000) & Cluster \\
\hline
464643791317333248	&	LW Cas	&	Suspected	&	44.33995	&	+60.68879	&	SAI 24	\\
1683014206596170240	&	CQ Dra	&	Confirmed	&	187.52767  &	+69.20088	&	HSC 1224	\\
5980480722594527872	&	[MMU2013] 355.12+03.82	&	Likely	&	259.63451	&	$-$30.92904 &	CWNU 170	\\
4058686267291546880	&	Gaia DR3 4058686267291546880	&	Suspected	&	261.93828 &	$-$29.86994	&	HSC 2962	\\
4066469336476655232	&	PPA J1808-2355	&	Suspected	&	272.07708	&	$-$23.92852	&	Collinder 367	\\
4155838500543591168	&	VPHASDR2J183044.6-100757.4	&	Suspected	&	277.68563 &	$-$10.13259 &	HSC 225	\\
2027037101299276928	&	SSTGLMC G062.9176+00.0981	&	Suspected	&	297.50614	&	+26.46056 &	HSC 509	\\
\end{tabular}
\end{center}
\end{table*}

A way out of this dilemma is the search for symbiotic stars among members of star clusters. 
Because all members of a star cluster are assumed to be born from one molecular cloud, 
they exhibit the same age \citep[and metallicity;][]{2021MNRAS.504..356D}. Up to now,
no symbiotic star has ever been confirmed in a Galactic globular cluster. 
\citet{2020MNRAS.496.3436B} analysed the possible astrophysical reasons and found that
most progenitors of these systems are destroyed through dynamical
interactions
in dense globular clusters before effectively becoming symbiotic stars. 
However, they should still be present in less dense clusters, but 
their overall rareness explains the absence.

With the availability of the $Gaia$ data set, it becomes more and more
important to discriminate between star clusters (remnants) and moving groups. The situation
becomes complicated when trying to prove if an aggregate was formed from one molecular cloud
and if the members are gravitationally bound \citep{2018ApJ...863...91F,2020A&A...638A.154K}.

In this letter, we concentrate on symbiotic stars being possible members of Galactic 
open clusters. 
Because of the $Gaia$ satellite mission \citep{2016A&A...595A...1G}, the analysis of open clusters 
made enormous progress \citep{2022Univ....8..111C}. Based on the $Gaia$ DR3 \citep{2023A&A...674A...1G}, several independent studies
analysed the members and cluster parameters of a significant number of 
aggregates \citep[e.g.][]{2023A&A...673A.114H,2024A&A...689A..18A}. We compared lists of
symbiotic stars and open clusters to find possible members.

\section*{Target Selection} \label{target_selection}

We took all established, likely, and suspected Galactic symbiotic stars 
from the recent version of the catalogue by \citet{2019RNAAS...3...28M}. 
We have not included the misclassified objects listed
in this database.
The matching was done via $Gaia$ IDs of objects from this list. 
All objects were cross-matched with 
the cluster member list and membership probabilities 
by \citet{2023A&A...673A.114H}.
This catalogue contains the parameters (age, reddening, and distance) of 7167 star clusters
including moving groups. They used the widely applied
Hierarchical Density-Based Spatial Clustering of Applications with Noise (HDBSCAN) algorithm \citep{McInnes2017}. It is well known that this method bears some limitations and flaws, mainly
when it is used as a numerical black box. Therefore, we utilised 
the positional, mean proper motion, and mean parallax information of the aggregates from
\citet{2023A&A...673A.114H} and all stars from the $Gaia$ DR3 \citep{2023A&A...674A...1G}.
The matching limits of \citet{2021A&A...645A..13P} were employed and 
additional members of open clusters were identified. This ensures the best possible
list of members using different approaches.
We found seven symbiotic star candidates as
possible members (Table \ref{table:canidates}).

\begin{figure}[!tb]
    \centering
    \includegraphics[width=\textwidth]{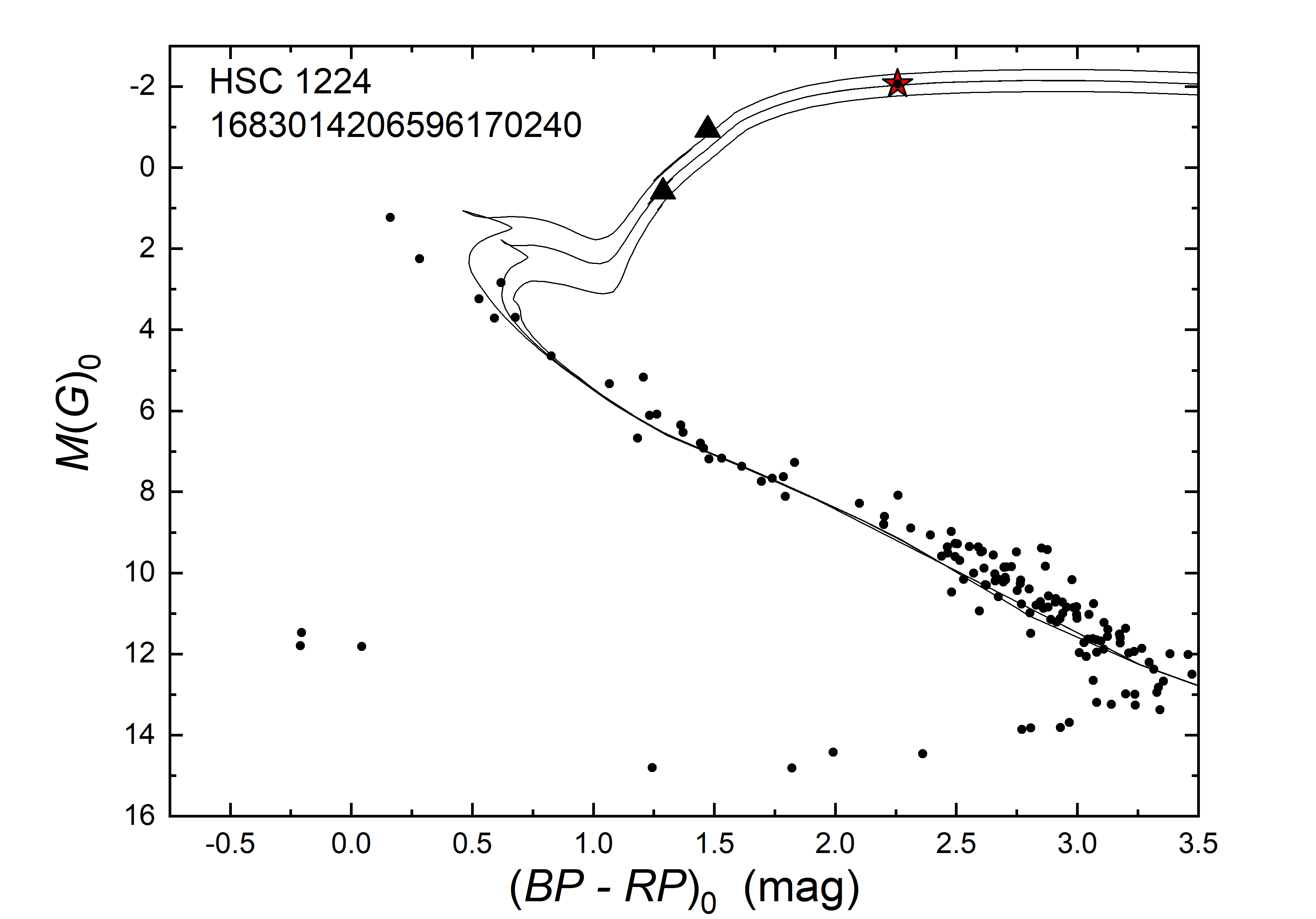}
    \caption{The only confirmed symbiotic star of our sample, $Gaia$ DR3 1683014206596170240
    and its host open cluster HSC 1244. The triangles denote HD 40325 and HD 73131, 
    two newly discovered members which support the derived age. The isochrones are from 
    \citet{2012MNRAS.427..127B} with $\log t$ values of 9.2, 9.4, and 9.6, respectively.}
    \label{fig:hsc}
\end{figure}

\begin{figure}[!tb]
    \centering
    \includegraphics[width=\textwidth]{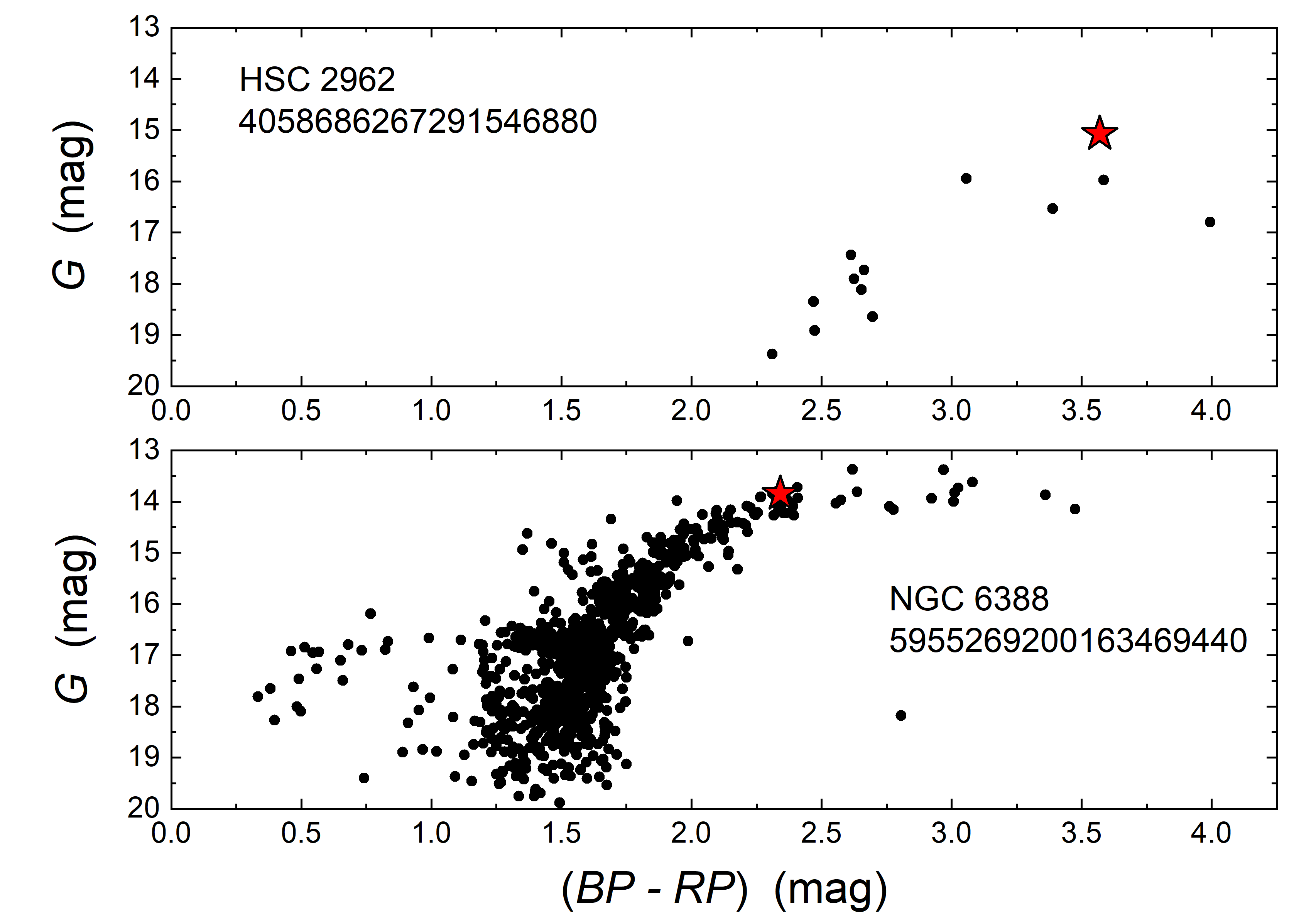}
    \caption{\textit{Upper}: $Gaia$ DR3 4058686267291546880 located in the cluster HSC 2962, classified as open cluster in \citet{2023A&A...673A.114H}. \textit{Lower}: The globular cluster NGC 6388 with the \textbf{candidate} symbiotic star $Gaia$ DR3 5955269200163469440 as a comparison. See text for a more detailed discussion.}
    \label{fig:gcls}
\end{figure}

\section*{Results} \label{results}

We must stress that it is still challenging to determine cluster parameters, 
although we already get a reasonable estimate of the distances 
from the $Gaia$ dataset \citep{2015A&A...582A..19N,2021MNRAS.504..356D}.\\

\textit{LW Cas} and \textit{SSTGLMC G062.9176+00.0981:} Here, we see  turn-off points. \citet{2023A&A...673A.114H,2024AJ....167...12C} list ages less than 10\,Myr
for both open clusters. Whereas SSTGLMC G062.9176+00.0981 seems to be a blue star, LW Cas is red. However, these stars can only be red giants if they have a significant selective reddening in the visual because their colours correspond to late-type objects. Another explanation for this unusual position for symbiotic stars would simply be misclassifications by previous studies. Especially LW Cas would become a Young Stellar Object (YSO) rather than a symbiotic star. Additionally, this star was classified as A0 III by \citet{1980AJ.....85...29C} and given the possibility of being an FU Ori star by \citet{1983MitVS...9..143W}, further making the classification as a symbiotic star debatable.\\

\textit{CQ Dra} This is our sample's only confirmed symbiotic star, first suspected by \citet{1989ApJ...345..489E} and confirmed via X-ray spectroscopy by \citet{2003MNRAS.346..855W}.  The possible host cluster, HSC 1224 is very close to the Sun (145\,pc) and listed with  an age estimation of 7.74\,$<$\,$\log t$\,$<$\,8.38 in \citet{2023A&A...673A.114H}.  As described in Sect. \ref{target_selection}, we searched for new members
of the host clusters and found two (HD 40325 and HD 73131) for HSC 1224. We used the reddening $E(BP-RP)$\,=\,0.072\,mag derived from CMDs using synthetic  $U$ magnitudes from BP/RP spectra \citep{2023A&A...674A..33G} and the corresponding distance to get the CMD shown in Figure \ref{fig:hsc}.  The newly found members are on the red giant branch, which matches their spectral classification of early K-type luminosity class III stars \citep{2012MNRAS.427..343M}. CQ Dra is located at the giant
branch with a $\log t$ of 9.4, respectively. Note the presence of three White Dwarfs within this open cluster also point towards an older age. The turn-off point is difficult to determine. It seems that there two Blue Stragglers are present, which is also not unusual for an old open cluster \citep{2021MNRAS.507.1699J}.\\

\textit{[MMU2013] 355.12+03.82:} This star was first reported as a candidate symbiotic star by \citet{2013MNRAS.432.3186M}. Based on their analysis, it shows emission in H I and maybe He I as well as absorption features of TiO and Na I D. The host open clusters
do not show any turn-off point and must be considered young. The Symbiotic candidate is located close to the main sequence. Only selective absorption can place it in the giant region of the CMDs. This can, of course, lead to the conclusion that the star was misclassified previously and could be a YSO.\\

\textit{$Gaia$} DR3 4058686267291546880: This star was reported as possible symbiotic star in \cite{2023A&A...674A..14R} by means of machine learning methods. It is located in HSC 2962, which is listed as an intermediate age (around 350\,Myr) open cluster with an extreme reddening of six magnitudes in the visual and a distance of about 5\,kpc \citep{2023A&A...673A.114H}. 
A comparison with the CMD of the  globular cluster NGC 6388
and the location of \textit{$Gaia$} DR3 5955269200163469440 \citep{2023A&A...674A..14R}
shows conspicuous similarities (Figure \ref{fig:gcls}). 
\citet{2023A&A...673A.114H} included NGC 6388
in their compilation and list an extremely high visual absorption of
3.3\,mag and an age estimation between 25\,Myr and 380\,Myr.
However, it is known that it has an age of at least 11\,Gyr and a reddening of only one magnitude at most \citep{2022A&A...659A.122C}. Also for
HSC 2962, the reddening values from different other sources \citep{2021MNRAS.508.1788A}
result in half the value given in \citet{2023A&A...673A.114H}. 
Taking realistic reddening, age and distance values, we conclude that $Gaia$ DR3 4058686267291546880
is on the red giant branch of the incorrectly classified globular cluster HSC 2962. However, due to the relative faintness and low number of the members reported by \cite{2023A&A...673A.114H}, this cluster is difficult to characterise in detail and thus may not exist. This idea is strengthened by the relatively low astrometric signal-to-noise ratio in \cite{2023A&A...673A.114H} who assign a value of $\sim 3.36$ to this quantitiy, while mentioning that all clusters with astrometric S/N below 5 are to be taken cautiously.\\

\textit{PPA J1808-2355}: The star is listed as a candidate symbiotic star in the Hong Kong/AAO/Strasbourg H$\alpha$ planetary nebula database \citep[HASH,][]{2016JPhCS.728c2008P}. It shows emission in the Balmer lines and also [NII]. It is also one of the stars that seem to be more on the (pre-)main sequence rather than on the giant branch. This again would only be possible due to either misclassification, extreme extinction or poor membership to the cluster. Further detailed analysis is needed.\\

\textit{VPHASDR2J183044.6-100757.4:} This is quite a particular case as seen in Figure \ref{fig:others}. The host open cluster can hardly be recognised by its CMD. It is 
a cloud of stars spreading over 1.5\,mag in colour and 2.5\,mag in apparent magnitude.
This is reflected by very different cluster parameters in the literature 
\citep{2023A&A...673A.114H,2024AJ....167...12C}.
VPHASDR2J183044.6-100757.4 is three magnitudes redder than the other stars, which would place it in the red giant region. The cluster is located in
the Galactic disc (Galactic latitude of $-$0.0696$\degr$) with a significant 
reddening. Therefore, getting more precise photometric data of fainter stars seems
to be difficult, but it is necessary to analyse this open cluster in more detail. The star was identified as symbiotic candidate by \citet{2019ApJS..240...21A} making use of machine learning methods based on infrared colours, specifically from 2MASS and WISE.

\subsection*{Check for YSO contamination}

A few of our candidates seem too close to the main sequence of their host clusters, making it difficult to decide from that alone if they are symbiotic stars (Figure \ref{fig:others}). The stars in question were checked for their spectral energy distributions (SED) in order to discern them from young stellar objects (YSO). We checked the Virtual Observatory SED Analyzer \citep[VOSA;][]{2008A&A...492..277B} for their characteristics in the infrared. Two out of three stars, LW Cas and PPA J1808-2355, show infrared excess, making it more likely for them to be YSOs rather than symbiotic binaries. In the case of [MMU2013] 355.12+03.82, VOSA could not detect any IR excess.

\section*{Conclusions and Outlook} \label{conclusions}

We presented a search for symbiotic stars and candidates that are members of Galactic open clusters.
Matching the newest lists resulted in seven good candidates. A closer look 
at the cluster parameters and CMDs revealed the shortcomings of 
the black-box algorithm in this research field. Specifically, there are problems to membership analysis using HDBSCAN which can be overconfident in detecting clusters, resulting in a number of false positives \citep{2023A&A...673A.114H}. Also, each machine learning method used in the literature comes with their own disadvantage in terms of how the data are processed and analysed. Additionally, we found that two symbiotic star candidates from the catalogue of \cite{2019RNAAS...3...28M} seem to be YSOs according to their SEDs. Our results show that symbiotic stars in open clusters can indeed be found and thus let us put more constraints on the age of these objects, given proper astrometry and cluster membership analysis. We see that especially for CQ Dra where we get to an age of $\log t \approx 9.4$.

We want to stress that proper motion and parallax measurements for extended objects and binaries might pose a problem within the $Gaia$ data sets. Initially, each object is treated as a single star. If the solution is unsatisfactory, several corrections caused by binarity were applied \citep{2023A&A...674A...9H}. This includes acceleration and orbital models for unresolved binaries with  either components that do not vary photometrically or one component that is always much brighter than the other. However, such corrections are always limited.\citet{2025A&A...695A..61M} mention an effect of the orbital period on the parallax measurements, making the parallax unreliable for shorter periods.
Another problem might be the parallax measurements of intrinsically extended objects, such as symbiotic stars with nebulae. A similar situation is found for  Planetary Nebulae for which \citet{2021A&A...656A..51G} discussed possible 
shortcomings. Therefore, several symbiotic stars as members of open clusters could be undetected. The upcoming fourth data release of the $Gaia$ consortium might resolve this issue.

\section*{Acknowledgements} 
This work was supported by the grant GA{\v C}R 23-07605S and
was carried out within the institutional support framework for the development of the
research organization of Masaryk University.
This research has made use of the SIMBAD database, operated at CDS, Strasbourg, France and of the WEBDA database, operated at the Department of Theoretical
Physics and Astrophysics of the Masaryk University.
This publication makes use of VOSA, developed under the Spanish Virtual Observatory (https://svo.cab.inta-csic.es) project funded by MCIN/AEI/10.13039/501100011033/ through grant PID2020-112949GB-I00. VOSA has been partially updated by using funding from the European Union's Horizon 2020 Research and Innovation Programme, under Grant Agreement no 776403 (EXOPLANETS-A).

\bibliographystyle{apalike}
\bibliography{paper_SySt}
\end{document}